\documentclass[10pt,aps,pra,twocolumn]{revtex4-2}
\usepackage{dcolumn}
\usepackage{times}
\usepackage{amsmath}
\usepackage{amsfonts}
\usepackage{amssymb}
\usepackage{bm}
\usepackage{bbm}
\usepackage{physics}
\usepackage{graphicx} 
\usepackage[table,xcdraw,dvipsnames]{xcolor}

\begin{document}
\title{High partial waves contribution in calculations of the polyvalent atoms}
\author{M. G. Kozlov$^{1, 2}$}

\affiliation{
$^1$ Petersburg Nuclear Physics Institute of NRC ``Kurchatov Institute'', Gatchina, Leningrad District, 188300, Russia \\
$^2$ Department of Physics, St. Petersburg Electrotechnical University “LETI”, Prof. Popov Street 5, St. Petersburg, 197376,  Russia
}
\begin{abstract}
High accuracy calculations of atomic properties require using long basis sets. In particular, it is necessary to include large number of partial waves and estimate truncation corrections. The convergence in partial waves is known to be rather slow, so calculations become very costly. We use valence perturbation theory [PRA \textbf{105}, 052805 (2022)] to calculate contribution of the high partial waves and estimate truncation corrections. These estimates may help to make assessment of theoretical error in atomic calculations more reliable.
\end{abstract}
\maketitle

\paragraph*{Introduction}
Complex many-electron atoms and ions are often used as sensitive instruments to study fundamental interactions and search for new physics beyond Standard model \cite{SBDJ18,Cong2025}. This requires developing more accurate methods of atomic calculations. Such calculations must accurately include relativistic and correlation effects, as well as main QED corrections. Recent studies \cite{SKBS24,KoYeKa24,KoKaDe2025,Cheung2025} demonstrated that the latter can be sufficiently accurately accounted for using effective QED potentials \cite{FlaGin05,TuBe13,STY13,Kaygorodov2021}. Thus, an accurate treatment of electronic correlations remains the most challenging part of the calculations of the many-electron atoms, in particular the ones with several valence electrons. Such calculations are done using basis sets of the one-electron atomic orbitals. Their accuracy depends on the length of the used basis sets and, in particular, on the number of the partial waves (PWs) - orbitals with different orbital angular momentum quantum number $l$. Including many PWs in the correlation calculation is very computationally expensive, so it is a common practice to use basis sets where the number of orbitals per PW rapidly decreases with $l$. This may lead to systematic underestimation of the contributions of the PWs with large $l$ and hampers accurate extrapolation to $l\to \infty$. 

In this article we suggest to calculate contribution of the high PWs using the valence many-body perturbation theory \cite{KTBM22}. This allows using longer basis sets with equal number of orbitals per PW and then make extrapolation to $l\to \infty$. As an example we calculate the spectrum of the neutral scandium. In this example we are not trying to perform a high accuracy calculation, where such an extrapolation would lead to an improvement of the agreement with the experimental spectrum. That would require a much more accurate treatment of the core-valence correlations and goes beyond the scope of the present study. We are also constraining our analysis to the energy spectrum, leaving aside other observables. One can expect that these corrections are decreasing faster for observables, which are described by the one-electron operators, than for the energy, which is described by the two-electron Hamiltonian. Otherwise, each such observable may have specific scaling law for the contributions of the high PWs and has to be studied separately.        

\paragraph*{Method}
In this paper we consider Scandium as a sufficiently complex atom with three valence electron over an argon-like core. Its ground state configuration is $[\mathrm{Ar}]3d 4s^2$. The wave function of the $3d$ electron strongly overlaps with outermost shells of the core, so the core-valence correlations are very important. They can be treated within CI+MBPT method (configuration interaction plus many-body perturbation theory) \cite{DFK96}, or using CI+all order method \cite{SKJJ09}. In these methods the core-valence correlations are treated either within MBPT method, or within all-order coupled cluster method. Here we are not trying to reach record accuracy and use simpler CI+MBPT approach. After that, the three-electron valence problem can be treated using conventional CI approach for the Dirac-Coulomb Hamiltonian. It is possible to include Breit and QED corrections, but we do not do it here. 

It is usually assumed that the three-electron valence problem can be treated relatively easily, as the respective CI matrix is typically not very large. However, when the high accuracy is necessary, one has to use very long basis sets and CI matrix becomes huge. For such cases in Ref.\ \cite{KTBM22} it was suggested to use CI+VPT (CI+valence perturbation theory) method where a part of the valence space is treated perturbatively. Here we use this method to account for the high PWs and estimate correction from the neglected ones.   

We started with the CI+MBPT calculation in the CI space, which included orbitals with $n=1,\dots,13$ and $l=0,\dots,3$ (basis set $[13spdf]$). In this basis set, the last PW is $f$ and it includes 10 orbitals with $n=4\dots 13$ for each total angular momentum $j=3\pm1/2$. After that, we made a serious of calculations adding each time one of the higher PWs from $g$ ($l=4$) to $m$ ($l=9$). The basis set for each of these PWs included 20 relativistic orbitals, 10 with $j=l-1/2$ and 10 with $j=l+1/2$. These orbitals were formed from B-splines using kinetic balance condition for the small component \cite{KozTup19}. On the final step, our calculation simultaneously included all PWs up to $l=9$. 

In the CI approach, adding 20 virtual orbitals for a single PW results in the huge increase of the size of the CI space. This makes each CI calculation very lengthy, even if we include only single (S) and double (D) excitations. The final calculation with all $6\times 20=120$ additional virtual orbitals is practically impossible within the CI approach. Because of that we treated high PW corrections within CI+VPT approach, suggested in Ref.\ \cite{KTBM22}. In this approach the high PW contribution to the energy of the atomic level is divided in two parts. The S excitations to these PWs are directly included into CI space, while the D excitations are treated by means of the MBPT and result in the corrections to the two-electron radial integrals for valence electrons \cite{KTBM22}. This allows to keep CI space relatively small and significantly speeds up calculations. 

Note that in the CI+MBPT method the core-valence correlations also lead to the corrections to the radial integrals for valence electrons. Here we simply add corrections to the radial integrals from the core-valence correlations and from the D excitations to the high PWs. At the same time, S excitations to high PWs are explicitly included in the CI space. 

\paragraph*{Results}
We calculated energies of the lowest 10 even and 10 odd states of Sc I, which group into 4 even and 3 odd multiplets. Results of the initial CI+MBPT calculation are listed in Table \ref{tab:E}. Here we use longer basis set $[13spdf]$, than in Ref.\ \cite{KTBM22}, but the accuracy of the present results is comparable to what was obtained there. 

\begin{table}[htb]
    \caption{Low-lying energy levels of Sc I (in cm$^{-1}$). CI+MBPT results are compared with experimental data from NIST \cite{NIST} and the average absolute difference is calculated as $|\Delta|_\mathrm{av} = \frac{1}{k}\sum_{i=1}^k |\Delta_i|$. In the last raw the total three-electron binding energy is given for the ground state in atomic units (Hartree).}
    \label{tab:E}
    \begin{tabular}{lcrrr}
    \hline\hline\\[-8pt]
    \multicolumn{1}{c}{Config.}&\multicolumn{1}{c}{Level}
    &\multicolumn{1}{c}{NIST}
    &\multicolumn{2}{c}{CI+MBPT}
    \\
    &&\multicolumn{1}{c}{$E_\mathrm{exper}$}&\multicolumn{1}{c}{$E_\mathrm{ther}$}&\multicolumn{1}{c}{$\Delta$}
   \\[2pt]
 $3d 4s^2$ & ${}^2D_{3/2}$   &$    0.0$&$     0.0$&$   0.0$\\
 $       $ & ${}^2D_{5/2}$   &$  168.3$&$   192.9$&$  24.6$\\[2pt]
 $3d^2 4s$ & ${}^4F_{3/2}$   &$11520.0$&$ 11072.0$&$-448.0$\\
 $       $ & ${}^4F_{5/2}$   &$11557.7$&$ 11188.5$&$-369.1$\\
 $       $ & ${}^4F_{7/2}$   &$11610.2$&$ 11352.3$&$-258.0$\\
 $       $ & ${}^4F_{9/2}$   &$11677.3$&$ 11561.7$&$-115.6$\\[2pt]
 $3d^2 4s$ & ${}^2F_{5/2}$   &$14926.1$&$ 14863.5$&$- 62.6$\\
 $       $ & ${}^2F_{7/2}$   &$15041.9$&$ 15169.2$&$-127.3$\\[2pt]
 $3d^2 4s$ & ${}^2D_{5/2}$   &$17012.8$&$ 17216.6$&$ 203.8$\\
 $       $ & ${}^2D_{3/2}$   &$17025.1$&$ 17457.6$&$ 432.5$\\[2pt]
 $3d4s4p $ & ${}^4F_{3/2}^o$ &$15672.6$&$ 15970.1$&$ 297.5$\\
 $       $ & ${}^4F_{5/2}^o$ &$15756.5$&$ 16079.5$&$ 323.0$\\
 $       $ & ${}^4F_{7/2}^o$ &$15881.7$&$ 16281.7$&$ 400.0$\\
 $       $ & ${}^4F_{9/2}^o$ &$16026.6$&$ 16489.7$&$ 463.1$\\[2pt]
 $3d4s4p $ & ${}^4D_{1/2}^o$ &$16009.7$&$ 16281.3$&$ 271.6$\\
 $       $ & ${}^4D_{3/2}^o$ &$16021.8$&$ 16348.1$&$ 326.3$\\
 $       $ & ${}^4D_{5/2}^o$ &$16141.0$&$ 16436.1$&$ 295.1$\\
 $       $ & ${}^4D_{7/2}^o$ &$16210.8$&$ 16548.7$&$ 337.9$\\[2pt]
 $3d4s4p $ & ${}^2D_{3/2}^o$ &$16022.7$&$ 16287.6$&$ 264.8$\\
 $       $ & ${}^2D_{5/2}^o$ &$16096.9$&$ 16288.2$&$ 191.3$\\[2pt]
 &\multicolumn{1}{c}{$|\Delta|_\mathrm{av}$}&
 &&{274.3}
\\
\multicolumn{2}{c}{$E_\mathrm{gs}$ (Hartree)}
                             &$1.621310$&$1.624646$&$0.003336$
\\
    \hline\hline
    \end{tabular}
\end{table}

\begin{table}[htb]
    \caption{Contribution of single excitations to the partial waves $l=4$~--~$7$ to the binding energies of low-lying multiplets of Sc I (in a.u.).}
    \label{tab:pwS}
    \begin{tabular}{lcrrrr}
    \hline\hline\\[-8pt]
    \multicolumn{1}{c}{Config.}&\multicolumn{1}{c}{Mult.}
    &\multicolumn{4}{c}{Partial wave}
    \\
    &&\multicolumn{1}{c}{$l=4$}&\multicolumn{1}{c}{$l=5$}&\multicolumn{1}{c}{$l=6$}&\multicolumn{1}{c}{$l=7$}
   \\[2pt]
 $3d 4s^2$ & ${}^2D_\mathrm{gs}$ & 6.39E-4  &  1.23E-4  & 2.14E-6  & 4.00E-8  \\  
 $3d^2 4s$ & ${}^4F$             & 9.13E-4  &  1.36E-4  & 2.05E-7  & 5.00E-9  \\  
 $3d^2 4s$ & ${}^2F$             & 1.34E-3  &  2.82E-4  & 1.72E-6  & 1.70E-7  \\  
 $3d^2 4s$ & ${}^2D$             & 1.41E-3  &  2.25E-4  & 6.60E-7  & 4.50E-8  \\  
 $3d4s4p $ & ${}^4F^o$           & 3.98E-4  &  5.84E-5  & 3.85E-6  & 3.75E-8  \\  
 $3d4s4p $ & ${}^4D^o$           & 3.74E-4  &  4.27E-5  & 4.23E-7  & 0.00E-0  \\  
 $3d4s4p $ & ${}^2D^o$           & 3.89E-4  &  6.62E-5  & 1.68E-6  & 1.50E-8  \\  
    \hline\hline
    \end{tabular}
\end{table}

\begin{table}[htb]
    \caption{Contribution of D excitations to the partial waves $l=4$~--~$9$ to the binding energies of lowest multiplets of Sc I (in a.u.).}
    \label{tab:pwD}
    \begin{tabular}{crrrrrr}
    \hline\hline\\[-8pt]
    \multicolumn{1}{c}{Mult.}
    &\multicolumn{6}{c}{Partial wave}
    \\
    &\multicolumn{1}{c}{$l=4$}&\multicolumn{1}{c}{$l=5$}&\multicolumn{1}{c}{$l=6$}
    &\multicolumn{1}{c}{$l=7$}&\multicolumn{1}{c}{$l=8$}&\multicolumn{1}{c}{$l=9$}
   \\[2pt]
 ${}^2D_\mathrm{gs}$ & 1.07E-4 & 4.78E-5 & 7.15E-5 & 3.27E-5 & 1.60E-5 & 7.54E-6 \\
 ${}^4F$             & 1.40E-4 & 2.85E-5 & 5.62E-5 & 1.99E-5 & 7.59E-6 & 3.80E-6 \\
 ${}^2F$             & 1.87E-4 & 5.42E-5 & 1.40E-4 & 6.17E-5 & 3.01E-5 & 1.54E-5 \\
 ${}^2D$             & 4.47E-4 & 1.19E-4 & 1.73E-4 & 7.90E-5 & 4.25E-5 & 2.29E-5 \\
 ${}^4F^o$           & 5.03E-6 & 1.45E-5 & 2.10E-5 & 9.87E-6 & 4.20E-6 & 2.22E-6 \\
 ${}^4D^o$           & 6.64E-6 & 3.50E-5 & 2.57E-5 & 9.51E-6 & 4.02E-6 & 1.75E-6 \\
 ${}^2D^o$           & 1.18E-5 & 2.17E-5 & 3.62E-5 & 1.63E-5 & 7.78E-6 & 4.34E-6 \\
    \hline\hline
    \end{tabular}
\end{table}

On the next step we sequentially add partial waves with $l$ from 4 to 9 and calculate level energies. The energy $E(S_{l_1},D_{l_2})$ corresponds to the calculation, where S excitations are includes for all PWs $l\le l_1$ and $D$ excitations are included for PWs $l\le l_2$. We define contribution of the S excitations to the PW $l$ as the difference
\begin{align} \label{eq:S_l}
    \Delta E_{S_l} &= 
        E(S_l,D_l)-E(S_{l-1},D_l)\,. 
\end{align}
We calculated these differences for $l\le 7$. Corrections to the levels of the same multiplet behave in a very similar manner, thus in Table \ref{tab:pwS} we present results for the values, averaged over each multiplet for brevity. As one can see, the corrections $\Delta E_{S_l}$ decrease very rapidly with $l$, so we did not calculate contributions from the PWs with $l=8,9$ and neglect them in the following discussion. 

The contribution of D excitations to the PW $l$ is defined as
\begin{align}\label{eq:D_l}
    \Delta E_{D_l} &= 
    \left\{
    \begin{array}{cc}
        E(S_{l-1},D_l)-E(S_{l-1},D_{l-1})\,, & l\le8\,,  \\
        E(S_{l=7},D_l)-E(S_{l=7},D_{l-1})\,, & l=9\,.
    \end{array}
    \right.
\end{align}
Note that definitions \eqref{eq:S_l} and \eqref{eq:D_l} guarantee that the sum $\Delta E_{S_l}+\Delta E_{D_l}\equiv \Delta E_l$ gives total contribution of the partial wave $l$ within the SD approximation. Corrections $\Delta E_{D_l}$ are listed in Table \ref{tab:pwD}.  

For the PW $l=4$, the S corrections dominate over D corrections. For the PW $l=5$, the S and D corrections are comparable, though S ones are still larger. Starting from the PW $l=6$, the D corrections dominate. For the PW $l=7$, the S corrections are two orders of magnitude smaller, than D ones. Therefore, we confirm again, that for PW with $l\ge 8$ the S corrections can be neglected.

\paragraph*{Scaling of corrections with $l$ and estimates of truncation errors}
In this section we try to analyze high PW corrections to the energies of atomic levels. As we already mentioned above, the corrections from the S excitations decrease much faster with $l$, than the corrections from the D excitations. Therefore, we either need to study these scalings separately, or sum them together and study scaling of the total corrections from the excitations to the PW $l$. \textit{Apriori} it is not clear, what is better. However, analysis of the data in Tables \ref{tab:pwS} -- \ref{tab:totSD} shows that total corrections have much smoother dependence on $l$. Thus, in the following we use the latter approach and show that it gives rather simple scaling rules. 

\begin{table}[htb]
    \caption{Total contribution of S and D excitations to the partial waves $l$ to the binding energies of lowest multiplets of Sc I 
    (in a.u.).}
    \label{tab:totSD}
    \begin{tabular}{crrrrrr}
    \hline\hline\\[-8pt]
    \multicolumn{1}{c}{Mult.}
    &\multicolumn{6}{c}{Partial wave}
    \\
    &\multicolumn{1}{c}{$l=4$}&\multicolumn{1}{c}{$l=5$}&\multicolumn{1}{c}{$l=6$}
    &\multicolumn{1}{c}{$l=7$}&\multicolumn{1}{c}{$l=8$}&\multicolumn{1}{c}{$l=9$}
   \\[2pt]
  ${}^2D_\mathrm{gs}$ & 7.46E-4 & 1.70E-4 & 7.37E-5 & 3.27E-5 & 1.60E-5 & 7.54E-6  \\  
  ${}^4F$             & 1.05E-3 & 1.64E-4 & 5.64E-5 & 1.99E-5 & 7.59E-6 & 3.80E-6  \\  
  ${}^2F$             & 1.52E-3 & 3.36E-4 & 1.42E-4 & 6.19E-5 & 3.01E-5 & 1.54E-5  \\  
  ${}^2D$             & 1.86E-3 & 3.45E-4 & 1.74E-4 & 7.90E-5 & 4.25E-5 & 2.29E-5  \\  
  ${}^4F^o$           & 4.04E-4 & 7.29E-5 & 2.49E-5 & 9.91E-6 & 4.20E-6 & 2.22E-6  \\  
  ${}^4D^o$           & 3.80E-4 & 7.77E-5 & 2.62E-5 & 9.51E-6 & 4.02E-6 & 1.75E-6  \\  
  ${}^2D^o$           & 4.01E-4 & 8.79E-5 & 3.79E-5 & 1.63E-5 & 7.78E-6 & 4.34E-6  \\  
    \hline\hline
    \end{tabular}
\end{table}

\begin{figure}
\centering
\includegraphics[width=1\linewidth]{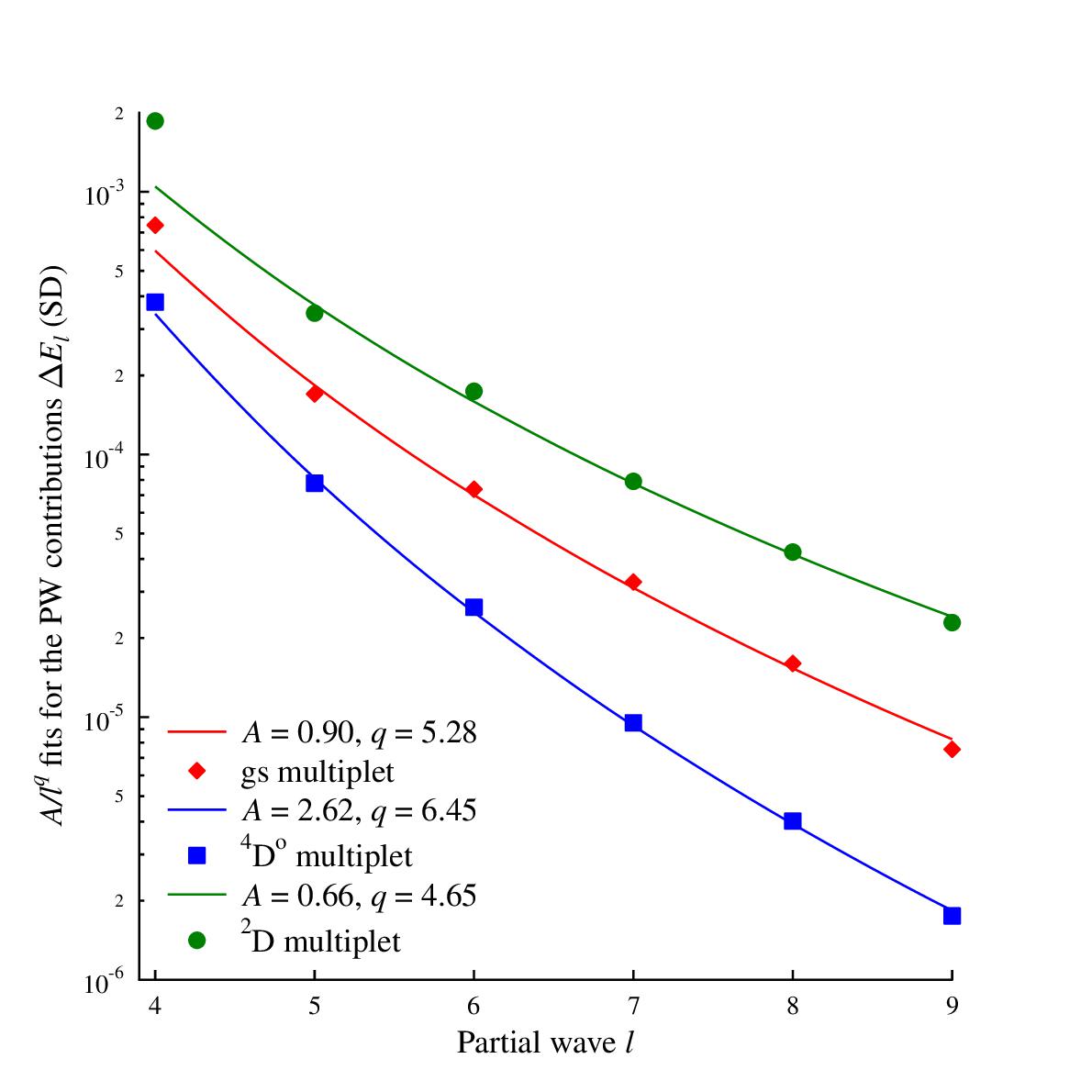}
    \caption{Corrections to the binding energies of Sc I for the ground state multiplet $^2D_\mathrm{gs}$ and multiplets ${}^2D$ and ${}^4D^o$ from S and D excitations to the partial wave $l$ and respective fits with functions ${A}/{L^q}$.}
    \label{fig:E_gs}
\end{figure}

The total corrections $\Delta E_l$ from the PW $l$ ($l=4,\dots, 9$) to the binding energies of the low-lying multiplets of Sc I are listed in Table \ref{tab:totSD}. Looking at different rows of this Table we see that corrections to the energies of different multiplets decrease differently with $l$. At the same time, it is easy to check, that along each row, $\log(\Delta E_l)/ \log l \approx$~const. Therefore, for each multiplet we use scaling function $A/l^q$, where $A$ and $q$ are fitting parameters. Results of such fits for three multiplets are shown in Fig.\ \ref{fig:E_gs}. 

One can see from Fig.\ \ref{fig:E_gs}, that, except for the contributions of the first PW $l=4$, all others are very well fitted by the functions $A/l^q$, with deviations being below 10\%. At the same time, the exponents for different multiplets vary significantly, from $q_\mathrm{min}=4.65$ to $q_\mathrm{max}=6.51$. The full list of parameters $A$ and $q$ is given in Table \ref{tab:A_q}.

\begin{table}[htb]
    \caption{Scaling and extrapolation of the contributions of SD excitations to high PWs to the binding energies of low-lying multiplets of Sc I. Function $f(L,q)$ is defined by Eq.\ \eqref{eq:Ratio}, $\Delta E_\mathrm{extrap}^{(9)}$ is the estimate of the contribution of PWs $l>9$ and $\Delta E_\mathrm{tot}^{(9)}=\sum_5^\infty \Delta E_l$ is the estimate of the contribution of all PWs $l>4$ (in a.u.).}
    \label{tab:A_q}
    \begin{tabular}{cdddccc}
    \hline\hline\\[-8pt]
    \multicolumn{1}{c}{Mult.}
    &\multicolumn{1}{c}{$A$} &\multicolumn{1}{c}{$q$}
    &\multicolumn{1}{c}{$f(9,q)$}  
    &\multicolumn{1}{c}{$\sum_5^9 \Delta E_l$} 
    &\multicolumn{1}{c}{$\Delta E_\mathrm{extrap}^{(9)}$}
    &\multicolumn{1}{c}{$\Delta E_\mathrm{tot}^{(9)}$}
   \\[2pt]
  ${}^2D_\mathrm{gs}$  & 0.897 & 5.281 & 1.65 & 3.00E-04 & 1.24E-05 & 3.13E-04  \\  
  ${}^4F$              & 6.135 & 6.511 & 1.19 & 2.52E-04 & 4.52E-06 & 2.57E-04  \\  
  ${}^2F$              & 1.677 & 5.261 & 1.66 & 5.85E-04 & 2.55E-05 & 6.11E-04  \\  
  ${}^2D$              & 0.658 & 4.648 & 2.01 & 6.63E-04 & 4.60E-05 & 7.09E-04  \\  
  ${}^4F^o$            & 1.123 & 5.988 & 1.36 & 1.14E-04 & 3.02E-06 & 1.17E-04  \\  
  ${}^4D^o$            & 2.619 & 6.450 & 1.21 & 1.19E-04 & 2.11E-06 & 1.21E-04  \\  
  ${}^2D^o$            & 0.385 & 5.183 & 1.70 & 1.54E-04 & 7.39E-06 & 1.62E-04  \\  
    \hline\hline
    \end{tabular}
\end{table}

\begin{figure}
\centering
\includegraphics[width=1\linewidth]{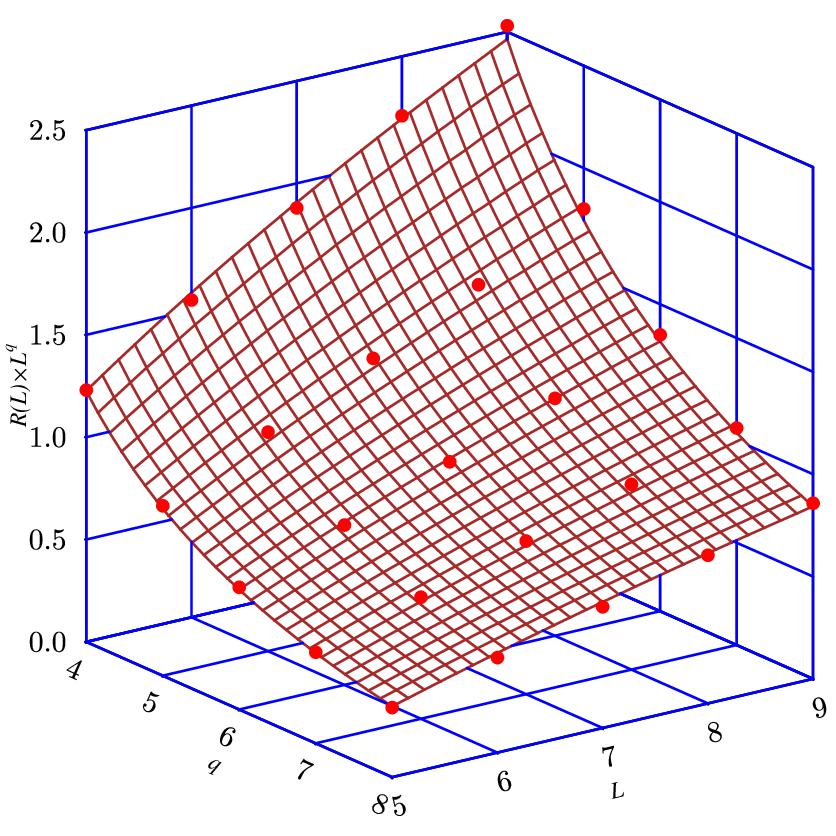}
    \caption{Function $R(L)\,{L^q}$ from Eq.\ \eqref{eq:Ratio} (red circles) and the fit with function \eqref{eq:Ratio_fit}. 
    }
    \label{fig:zeta_3D}
\end{figure}

Large differences in scalings for different levels result in rather complex behavior of the corrections to the transition frequencies between levels. Thus, we continue with the estimation of the error from the truncation of the PW expansion for the binding energies, rather than for the transition frequencies. 

If the corrections from the high PW contributions are described by the function $A/l^q$, then the truncation error is given by the residue:
\begin{multline}\label{eq:Res_def}
    AR(L) \equiv E(S_\infty,D_\infty)-E(S_{L},D_L)
    \\ \approx A\left( \zeta(q) -\sum_{l=1}^L \frac{1}{l^q} 
    \right)\,,
\end{multline}
where $\zeta(q)=\sum_{l=1}^\infty l^{-q}$ is Riemann $\zeta$-function. It is relatively common to estimate truncation error in terms of the last accounted contribution. Thus, let us look at the ratio:  
\begin{align}
\label{eq:Ratio}
    f(L,q) \equiv 
    \frac{A R(L)}{A/L^{q}} =R(L) L^{q}
    = L^q\! \left( \zeta(q) -\sum_{l=1}^L \frac{1}{l^q} 
    \right)\!.
\end{align}
We see that this ratio does not depend on the fitting parameter $A$, but it does depend on both $L$ and $q$. Figure \ref{fig:zeta_3D} shows ratio \eqref{eq:Ratio} for different values of $L$ and $q$. It is clear from this plot that there is no universal relation between the last contribution and the truncation error. Firstly, the former decreases faster with $L$ than the latter. Secondly, the ratio is larger for smaller values of $q$. In Fig.\ \ref{fig:zeta_3D} we also plot the fit of the ratio \eqref{eq:Ratio} with the function  
\begin{align}
  \label{eq:Ratio_fit}
  \begin{array}{ll}
       \multicolumn{2}{c}{f(L,q) = \frac{F_0+F_1 (L-7)}{q^{a_0+a_1/L^2}}\,;}  \\
       F_0 = 19.06\,, & F_1 = 1.483\,;\\ 
       a_0 = 1.445\,, & a_1 = 10.31\,.
  \end{array}
\end{align}

If corrections to the energy of atomic level follow the $A/L^q$ scaling, we can use the correction from the last PW and either Eq.\ \eqref{eq:Ratio}, or Eq.\ \eqref{eq:Ratio_fit} to estimate contribution from neglected PWs with $l>L$. This is done in Table \ref{tab:A_q}. Then, this estimate can be added to the result of the calculation to get extrapolated value of the energy (see the last column in Table \ref{tab:A_q}). 

The accuracy of such extrapolation to $L\to \infty$ depends on how reliably we know the scaling exponents $q$. Finding their values may require calculations for sufficiently many PWs. In order to test the consistency and accuracy of such extrapolations we performed calculations of the exponents $q$ and extrapolated high PW corrections to the energies $\Delta E_\mathrm{tot}^{(L)}$ from the truncated arrays of data for $L= 8$ and $L=7$ respectively. These results are summarized in Table \ref{tab:comparisson}. Results of all three extrapolations are different, but these differences are few times smaller than the last accounted correction $\Delta E_L$. We calculated the following ratios: 
\begin{align}
    \label{eq:compare}
    \chi_L &=(E_\mathrm{tot}^{(9)}-E_\mathrm{tot}^{(L)})/\Delta E_L\,, 
\end{align}
for $L=7,8$ and listed them in Table \ref{tab:comparisson}. One can see, that the values of $\chi_8$ and $\chi_7$ are similar and all of them are small, $|\chi_L|<0.3$.

\begin{table}[htb]
    \caption{Extrapolation to $L\to \infty$ from $L=9$, $L=8$, and $L=7$ for energies (in a.u.). Parameters $\chi_L$ characterize the differences between extrapolations, see Eq.\ \eqref{eq:compare}. }
    \label{tab:comparisson}
    \begin{tabular}{crrdrd}
    \hline\hline\\[-8pt]
    &\multicolumn{1}{c}{$L=9$} 
    &\multicolumn{2}{c}{$L=8$} 
    &\multicolumn{2}{c}{$L=7$} \\
    \multicolumn{1}{c}{Mult.}
    &\multicolumn{1}{c}{$\Delta E_\mathrm{tot}^{(9)}$}
    &\multicolumn{1}{c}{$\Delta E_\mathrm{tot}^{(8)}$}  
    &\multicolumn{1}{c}{$\chi_8$}
    &\multicolumn{1}{c}{$\Delta E_\mathrm{tot}^{(7)}$}
    &\multicolumn{1}{c}{$\chi_7$}
   \\[2pt]
  ${}^2D_\mathrm{gs}$  & 3.13E-04 & 3.17E-04 & -0.28 & 3.21E-04 & -0.25  \\
  ${}^4F$              & 2.57E-04 & 2.56E-04 &  0.09 & 2.59E-04 & -0.10  \\
  ${}^2F$              & 6.11E-04 & 6.15E-04 & -0.12 & 6.20E-04 & -0.15  \\
  ${}^2D$              & 7.09E-04 & 7.18E-04 & -0.21 & 7.27E-04 & -0.23  \\
  ${}^4F^o$            & 1.17E-04 & 1.17E-04 &  0.10 & 1.18E-04 & -0.04  \\
  ${}^4D^o$            & 1.21E-04 & 1.22E-04 & -0.11 & 1.22E-04 & -0.08  \\
  ${}^2D^o$            & 1.62E-04 & 1.61E-04 &  0.04 & 1.64E-04 & -0.12  \\
    \hline\hline
    \end{tabular}
\end{table}

\begin{table}[htb]
    \caption{Extrapolation to $L\to \infty$ from $L=9$, $L=8$, and $L=7$ for transition frequencies from the ground state (in cm$^{-1}$). Relative deviations between extrapolations $\chi_i$ are defined by Eq.\ \eqref{eq:compare} with energies substituted with the respective frequencies.}
    \label{tab:transition}
    \begin{tabular}{cddddd}
    \hline\hline\\[-8pt]
    &\multicolumn{1}{c}{$L=9$} 
    &\multicolumn{2}{c}{$L=8$} 
    &\multicolumn{2}{c}{$L=7$} \\
    \multicolumn{1}{c}{Mult.}
    &\multicolumn{1}{c}{$\Delta \omega_\mathrm{tot}^{(9)}$}
    &\multicolumn{1}{c}{$\Delta \omega_\mathrm{tot}^{(8)}$}  
    &\multicolumn{1}{c}{$\chi_8$}
    &\multicolumn{1}{c}{$\Delta \omega_\mathrm{tot}^{(7)}$}
    &\multicolumn{1}{c}{$\chi_7$}
   \\[2pt]
  ${}^4F$              &  12.33 &  13.44 & -0.61 &  13.69 & -0.48 \\
  ${}^2F$              & -65.41 & -65.27 &  0.05 & -65.69 & -0.04 \\
  ${}^2D$              & -86.94 & -87.94 & -0.17 & -89.13 & -0.22 \\
  ${}^4F^o$            &  42.93 &  43.99 & -0.41 &  44.65 & -0.34 \\
  ${}^4D^o$            &  42.04 &  42.91 & -0.33 &  43.67 & -0.32 \\
  ${}^2D^o$            &  33.17 &  34.20 & -0.58 &  34.55 & -0.39 \\
    \hline\hline
    \end{tabular}
\end{table}

Figures \ref{fig:E_gs} and \ref{fig:zeta_3D} show, that if we stopped our calculations at $L=5$, or $L=6$, our truncation errors would be smaller, than the last calculated corrections $\Delta E_L$. However, if we stopped our calculation at $L=7$, or higher, our truncation errors would be generally larger, than the last calculated corrections $\Delta E_L$. In this case, it would be safer to estimate the error as twice the size of the last correction. On the other hand, if we make extrapolation to $L\to\infty$, we can estimate our theoretical error to be about one third of the last correction, or less. This means that extrapolation allows us to reduce the theoretical error by three to six times.

Now let us briefly discuss the high PW corrections to the transition frequencies from the ground state, $\hbar\omega_i=E_i-E_\mathrm{gs}$. Average corrections to the transition frequencies to the levels of the low-lying multiplets can be found using the data from Tables \ref{tab:totSD} -- \ref{tab:comparisson}, see Table \ref{tab:transition}. We see that these corrections for the transitions to levels of the configurations $3d^2 4s$ and $3d4s4p$ have different signs. Also, the differences between extrapolations from different $L$ are noticeably larger, than for the energies. This is not surprising because corrections to the transition frequencies depend on the differences between the corrections to the upper and lower levels, which have different scalings with $L$. Never the less, we see that the differences between extrapolations from different $L$ is smaller than the last accounted correction, see the values of the ratios $\chi_8$ and $\chi_7$ in Table \ref{tab:transition}, which are defined by Eq.\ \eqref{eq:compare} with energies substituted with the respective frequencies.   

\paragraph*{Summary}
In this paper we studied corrections to the binding energies of the low-lying levels of Sc I from the virtual orbitals with angular momenta $l\ge 4$. We found that starting from $l=5$, these corrections scale as $A/l^q$, with exponents $q$ varying from $q_\mathrm{min}\approx 4.7$ to $q_\mathrm{max}\approx 6.5$. These findings are consistent with findings in Refs.\ \cite{Skripnikov2021,Athanasakis2025} for Ra$^+$ ion where similar scalings with $q\approx 6$ were also observed. Such scalings allow estimation of the truncation corrections from the PWs $l>L$, which are not included in the calculation. If the scaling exponent is known, the size of these corrections can be related to the contribution of the last included PW $l=L$. The accuracy of such extrapolation critically depends on the accuracy of the calculation of the last contribution $\Delta E_L$ and on the scaling exponent $q$. To this end it is important to make calculations with the sufficiently saturated basis sets for high PW. 

A relatively common practice to use basis sets, where the number of basis orbitals per PW rapidly decreases with $l$, may lead to underestimation of $\Delta E_L$ with simultaneous overestimation of $q$. Both these errors lead to underestimation of the truncation correction. In Ref.\ \cite{Kaygorodov2019} the ground state energy of the Be-like Ne was calculated using long basis sets including PW up to $l=10$ and the number of orbitals per PW $k$ from 10 to 30. The fits of their data for $k=10$ and for $k=30$ give exponents $q=3.83$ and $q=3.59$, respectively. Thus, our present calculation with $k=10$ may still underestimate scaling exponent by approximately 10\%.  

Another question we tried to address in this study is how many PWs must be included in the calculation in order to make reliable extrapolation to $L\to \infty$. We found out that smooth scaling of the corrections start from $L=5$, which is again consistent with finding in Refs.\ \cite{Skripnikov2021,Athanasakis2025}. Assuming that we need at least three contributions to determine scaling exponent $q$, we come to the conclusion that calculation must include at least all PWs up to $L=7$. Comparing extrapolations to the infinity from $L=7,8,\,\mathrm{and}\,\, 9$, we see that they are consistent with each other. The absolute values of the differences between these extrapolations are less than 30\% of the last calculated contribution. Thus, we can assume that extrapolation error for the valence binding energy is about 30\% of the last contribution.

In many practical applications we are interested in transition frequencies between atomic levels, rather than in their binding energies. Extrapolation to $L\to \infty$ in such cases is more difficult. Indeed, we see, that scaling exponents significantly differ from one atomic level to another. Therefore, extrapolation for the transition frequency requires not two parameters, but four. Truncation error and uncertainty of the extrapolation is higher. Still, for the transitions from the ground state studied here the error of the extrapolation to $L\to\infty$ is probably smaller than the last accounted correction. 

Note that the present study is based on the calculations for Sc I. We chose this atom because it has three valence electrons, which occupy $4s$, $4p$, and $3d$ orbitals. This distinguishes Sc from the more studied cases of one- and two-electron atoms, where electrons typically occupy $s$ and $p$ orbitals. It is still to be checked if our conclusions are applicable for other atoms, in particular, for atoms with open $f$ shells. 

\paragraph*{Acknowledgments}
The author is grateful to Leonid Skripnikov, Dmitry Glazov, Ilya Tutitzyn, Vladimir Dzuba, Alexey Malyshev, and Egor Lazarev for very helpful discussions.
This work was supported by the Foundation for the Advancement of Theoretical
Physics and Mathematics ``BASIS''.

\bibliography{./refs}

@Article{SBDJ18,
  author    = {M. S. Safronova and D. Budker and D. DeMille and D. F. {Jackson Kimball} and A. Derevianko and C. W. Clark},
  title     = "{Search for New Physics with Atoms and Molecules}",
  eid       = {025008},
  eprint    = {1710.01833},
  archiveprefix = {arXiv},
  volume    = {90},
  journal   = {Rev. Mod. Phys.},
  year      = {2018},
}

@Article{Cong2025,
  author        = {Cong, Lei and Ji, Wei and Fadeev, Pavel and Ficek, Filip and Jiang, Min and Flambaum, Victor V. and Guan, Haosen and Jackson Kimball, Derek F. and Kozlov, Mikhail G. and Stadnik, Yevgeny V. and Budker, Dmitry},
  date          = {2025-06},
  journaltitle  = {Rev. Mod. Phys.},
  title         = {Spin-dependent exotic interactions},
  doi           = {10.1103/RevModPhys.97.025005},
  eprint        = {2408.15691},
  issue         = {2},
  pages         = {025005},
  volume        = {97},
  archiveprefix = {arXiv},
  journal       = {Rev. Mod. Phys.},
  numpages      = {86},
  year          = {2025},
}

@Article{SKBS24,
  author   = {Chintan Shah and Steffen Kühn and Sonja Bernitt and Ren{\'{e}} Steinbrügge and Moto Togawa and Lukas Berger and Jens Buck and Moritz Hoesch and Jörn Seltmann and Mikhail G. Kozlov and Sergey G. Porsev and Ming Feng Gu and F. {Scott Porter} and Thomas Pfeifer and Maurice A. Leutenegger and Charles Cheung and Marianna S. Safronova and Jos{\'{e}} R. Crespo L{\'{o}}pez-Urrutia},
  title    = {{Natural-linewidth measurements of the 3C and 3D soft-x-ray transitions in Ni XIX}},
  doi      = {10.1103/PhysRevA.109.063108},
  pages    = {063108},
  volume   = {109},
  journal  = {Phys. Rev. A},
  year     = {2024},
}

@Article{KoYeKa24,
  author        = {M. G. Kozlov and V. A. Yerokhin and M. Y. Kaygorodov and E. V. Tryapitsyna},
  year          = {2024},
  journal       = {Phys. Rev. A},
  title         = {{QED calculations of the E1 transition amplitude in neonlike iron and nickel}},
  doi           = {10.1103/PhysRevA.110.062805},
  eprint        = {2410.02489},
  pages         = {062805},
  volume        = {110},
  archiveprefix = {arXiv},
}

@Article{KoKaDe2025,
  author        = {Kozlov, M. G. and Kaygorodov, M. Y. and Demidov, Yu. A. and Yerokhin, V. A.},
  date          = {2025-02},
  title         = {Self-energy correction to the E1 transition amplitudes in hydrogenlike ions},
  doi           = {10.1103/physreva.111.022816},
  eprint        = {2412.01231},
  issn          = {2469-9934},
  number        = {2},
  pages         = {022816},
  volume        = {111},
  archiveprefix = {arXiv},
  journal       = {Physical Review A},
  publisher     = {American Physical Society (APS)},
  year          = {2025},
}

@Article{Cheung2025,
  author       = {Cheung, Charles and Porsev, Sergey G. and Filin, Dmytro and Safronova, Marianna S. and Wehrheim, Malte and Spieß, Lukas J. and Chen, Shuying and Wilzewski, Alexander and López-Urrutia, José R. Crespo and Schmidt, Piet O.},
  journal      = {Physical Review Letters},
  title        = "{Finding the Ultranarrow $^3$P$_2$ → $^3$P$_0$ Electric Quadrupole Transition in Ni$^{12+}$ Ion for an Optical Clock}",
  year         = {2025},
  number       = {9},
  volume       = {135},
  date         = {2025-08},
  doi          = {10.1103/flwf-c2m1},
  eprint       = {2502.05386},
  eprintclass  = {physics.atom-ph},
  eprinttype   = {arXiv},
  journaltitle = {Physical Review Letters},
  publisher    = {American Physical Society (APS)},
}

@Article{FlaGin05,
  author    = {{Flambaum}, V. V. and {Ginges}, J. S. M.},
  title     = {{Radiative potential and calculations of QED radiative corrections to energy levels and electromagnetic amplitudes in many-electron atoms}},
  doi       = {10.1103/PhysRevA.72.052115},
  number    = {5},
  pages     = {052115},
  volume    = {72},
  journal   = {Phys. Rev. A},
  month     = {nov},
  year      = {2005},
}

@Article{TuBe13,
  author    = {I. I. Tupitsyn and E. V. Berseneva},
  title     = {A Single-Particle Nonlocal Potential for Taking into Account Quantum-Electrodynamic Corrections in Calculations of the Electronic Structure of Atoms},
  pages     = {682},
  volume    = {114},
  journal   = {Opt. Spectrosc.},
  year      = {2013},
}

@Article{STY13,
  author   = {Shabaev, V. M. and Tupitsyn, I. I. and Yerokhin, V. A.},
  title    = "{Model operator approach to the Lamb shift calculations in relativistic many-electron atoms}",
  doi      = {10.1103/PhysRevA.88.012513},
  issue    = {1},
  pages    = {012513},
  volume   = {88},
  journal  = {Phys. Rev. A},
  month    = {Jul},
  year     = {2013},
}

@Article{Kaygorodov2021,
  author       = {Kaygorodov, M. Y. and Skripnikov, L. V. and Tupitsyn, I. I. and Eliav, E. and Kozhedub, Y. S. and Malyshev, A. V. and Oleynichenko, A. V. and Shabaev, V. M. and Titov, A. V. and Zaitsevskii, A. V.},
  journal      = {Physical Review A},
  title        = {Electron affinity of oganesson},
  year         = {2021},
  issn         = {2469-9934},
  number       = {1},
  pages        = {012819},
  volume       = {104},
  date         = {2021-07},
  doi          = {10.1103/physreva.104.012819},
  journaltitle = {Physical Review A},
  publisher    = {American Physical Society (APS)},
}

@Article{KTBM22,
  author        = {Kozlov, M. G. and Tupitsyn, I. I. and Bondarev, A. I. and Mironova, D. V.},
  title         = {Combination of perturbation theory with the configuration-interaction method},
  doi           = {10.1103/PhysRevA.105.052805},
  eprint        = {2202.02026},
  issue         = {5},
  pages         = {052805},
  volume        = {105},
  archiveprefix = {arXiv},
  journal       = {Phys. Rev. A},
  numpages      = {10},
  year          = {2022},
}

@Article{DFK96,
  author    = {V. A. Dzuba and V. V. Flambaum and M. G. Kozlov},
  title     = {Combination of the many-body perturbation theory with the configuration-interaction method},
  journal   = {Phys. Rev. A},
  year      = {1996},
  volume    = {54},
  number    = {5},
  pages     = {3948--3959},
  month     = {nov},
  doi       = {10.1103/physreva.54.3948},
  publisher = {American Physical Society ({APS})},
}

@Article{SKJJ09,
  author   = {M S Safronova and M G Kozlov and W R Johnson and D Jiang},
  title    = {Development of a configuration-interaction + all-order method for atomic calculations},
  journal  = {Phys. Rev. A},
  year     = {2009},
  volume   = {80},
  pages    = {012516},
  doi      = {10.1103/PhysRevA.80.012516},
  eprint   = {0905.2578},
  archiveprefix = {arXiv},
}

@Article{KozTup19,
  author  = {Mikhail Kozlov and Ilya Tupitsyn},
  title   = {Mixed Basis Sets for Atomic Calculations},
  journal = {Atoms},
  year    = {2019},
  volume  = {7},
  number  = {3},
  pages   = {92},
  doi     = {10.3390/atoms7030092},
}

@Misc{NIST,
  author    = {A. Kramida and Yu. Ralchenko and J. Reader and {NIST ASD Team}},
  title     = "{NIST Atomic Spectra Database}",
  url       = {http://physics.nist.gov/PhysRefData/ASD/index.html},
  year      = {2016},
}

@Article{Skripnikov2021,
  author       = {Skripnikov, Leonid V.},
  journal      = {The Journal of Chemical Physics},
  title        = {{Approaching meV level for transition energies in the radium monofluoride molecule RaF and radium cation Ra+ by including quantum-electrodynamics effects}},
  year         = {2021},
  number       = {20},
  volume       = {154},
  eid          = {201101},
  date         = {2021-05},
  doi          = {10.1063/5.0053659},
  journaltitle = {The Journal of Chemical Physics},
  publisher    = {AIP Publishing},
}

@Article{Athanasakis2025,
  author       = {Athanasakis-Kaklamanakis, M. and Wilkins, S. G. and Skripnikov, L. V. and others},
  journal      = {Nature Communications},
  title        = {Electron correlation and relativistic effects in the excited states of radium monofluoride},
  year         = {2025},
  issn         = {2041-1723},
  number       = {1},
  volume       = {16},
  eid          = {2139},
  date         = {2025-03},
  doi          = {10.1038/s41467-025-55977-w},
  journaltitle = {Nature Communications},
  publisher    = {Springer Science and Business Media LLC},
}

@Article{Kaygorodov2019,
  author       = {Kaygorodov, M. Y. and Kozhedub, Y. S. and Tupitsyn, I. I. and Malyshev, A. V. and Glazov, D. A. and Plunien, G. and Shabaev, V. M.},
  journal      = {Physical Review A},
  title        = {{Relativistic calculations of the ground and inner- L-shell excited energy levels of berylliumlike ions}},
  year         = {2019},
  issn         = {2469-9934},
  number       = {3},
  pages        = {032505},
  volume       = {99},
  date         = {2019-03},
  doi          = {10.1103/physreva.99.032505},
  journaltitle = {Physical Review A},
  publisher    = {American Physical Society (APS)},
}

\end{document}